\documentclass[twocolumn,trackchanges]{aastex6}

\usepackage{amssymb,amsmath}
\usepackage{color}

\newcommand{\halpha}{H$\alpha$}
\newcommand{\nh}{$N_{\rm H}$}

\received{}
\revised{}
\accepted{}
\published{}

\shorttitle{X-ray Absorbed Type 1 AGN} 
\shortauthors{Shimizu T. T. \emph{et\,al.}}

\begin{document}

\title{BAT AGN Spectroscopic Survey--VIII. Type 1 AGN With Massive Absorbing Columns}
\author{T. Taro Shimizu$^{1,\dagger}$, Richard I. Davies$^{1}$, Michael Koss$^{2,3}$, Claudio Ricci$^{4,5,6}$, Isabella Lamperti$^{3,9}$, Kyuseok Oh$^{3}$, Kevin Schawinski$^{3}$, Benny Trakhtenbrot$^{3}$, Leonard Burtscher$^{7}$, Reinhard Genzel$^{1}$, Ming-yi Lin$^{1}$, Dieter Lutz$^{1}$, David Rosario$^{8}$, Eckhard Sturm$^{1}$, Linda Tacconi$^{1}$}

\affiliation{$^{1}$Max-Planck-Institut f\"{u}r extraterrestrische Physik, Postfach 1312, 85741, Garching, Germany \\
$^{2}$Eureka Scientific Inc., 2452 Delmer St. Suite 100, Oakland, CA 94602, USA\\
$^{3}$Institute for Astronomy, Department of Physics, ETH Zurich, Wolfgang-Pauli-Strasse 27, CH-8093 Zurich, Switzerland\\
$^{4}$Instituto de Astrofisica, Pontificia Universidad Cat\'olica de Chile, Vicu\~{n}a Mackenna 4860, Santiago, Chile\\
$^{5}$Kavli Institute for Astronomy and Astrophysics, Peking University, Beijing 100871, China\\
$^{6}$Chinese Academy of Sciences South America Center for Astronomy and China-Chile Joint Center for Astronomy, Camino El Observatorio 1515, Las Condes, Santiago, Chile\\
$^{7}$Sterrewacht Leiden, Universiteit Leiden, Niels-Bohr-Weg 2, 2300 CA Leiden, The Netherlands\\
$^{8}$Department of Physics, Durham University, South Road, DH1 3LE, Durham, UK\\
$^{9}$Astrophysics Group, Department of Physics and Astronomy, University College London, 132 Hampstead Road, London NW1 2PS, UK\\
}

\email{$^{\dagger}$shimizu@mpe.mpg.de}

\begin{abstract}
We explore the relationship between X-ray absorption and optical obscuration within the BAT AGN Spectroscopic Survey (BASS) which has been collecting and analyzing the optical and X-ray spectra for 641 hard X-ray selected ($E>14$ keV) active galactic nuclei (AGN).  We use the deviation from a linear broad \halpha-to-X-ray relationship as an estimate of the maximum optical obscuration towards the broad line region and compare the $A_{\rm V}$ to the hydrogen column densities (\nh) found through systematic modeling of their X-ray spectra. We find that the inferred columns implied by  $A_{\rm V}$ towards the broad line region (BLR) are often orders of magnitude less than the columns measured towards the X-ray emitting region indicating a small scale origin for the X-ray absorbing gas. After removing 30\% of Sy 1.9s that potentially have been misclassified due to outflows, we find that 86\% (164/190) of the Type 1 population (Sy 1--1.9) are X-ray unabsorbed as expected based on a single obscuring structure. However, 14\% (26/190), of which 70\% (18/26) are classified as Sy 1.9, are X-ray absorbed, suggesting the broad line region itself is providing extra obscuration towards the X-ray corona. The fraction of X-ray absorbed Type 1 AGN remains relatively constant with AGN luminosity and Eddington ratio, indicating a stable broad line region covering fraction.
\end{abstract}


\section{Introduction}\label{sec:intro}
The unified model of active galactic nuclei \citep[AGN]{Antonucci:1993os, Urry:1995il} attributes the differences between Type 1 and Type 2 AGN to changes in the orientation of our line of sight with respect to a large obscuring structure encircling the AGN. The common model for the obscuring structure is a torus consisting of cold gas and dust with recent work strongly suggesting a clumpy distribution \citep[e.g.][]{Alonso-Herrero:2003hc, Nenkova:2008dq, Nikutta:2009dw, Honig:2010lr, Mor:2012fj, Markowitz:2014aa}. Near the central supermassive black hole (SMBH) and within the inner radius of the torus is thought to be the broad line region (BLR) which consists of high velocity clouds that produce the typical broad emission lines (full width at half maximum, FWHM $\gtrsim2000$ km s$^{-1}$) seen in Type 1 AGN.  Type 1 AGN, therefore, are observed at angles above the torus with a direct view of the BLR while Type 2 AGN, which only show narrow emission lines, are observed through the torus that obscures our view of the BLR \citep[for a complete review of the unified model see][]{Netzer:2015aa}. 

A completely independent method for differentiating between Type 1 and Type 2 AGN is by directly measuring the neutral hydrogen column density (\nh) from X-ray spectra. Hard X-ray emission ($>10$ keV) is ubiquitous in all but the most highly obscured AGN \citep[Compton thick; \nh $\gtrsim10^{24}$ cm$^{-2}$; e.g.][]{Koss:2016eh} and is thought to originate from a compact corona near the SMBH \citep[e.g.][]{Haardt:1994xl}. Intervening neutral gas along our line of sight absorbs X-rays up to an energy cutoff that is dependent on the column of gas. 

Therefore both UV-optical and X-ray observations are useful tracers of the dust and gas distribution around AGN and any relationships that exist between obscuration/absorption and AGN properties can provide insight into how the AGN controls and effects the environment within which it lives. Under the simple picture of a static dusty torus around an AGN, both optical and X-ray measurements of the gas column density should agree. To a large extent this is true as many studies find that Type 1 AGN show little to no X-ray absorption while most Type 2 AGN are X-ray absorbed with \nh{} $\gtrsim10^{22}$ cm$^{-2}$ \citep[e.g.][]{Smith:1996aa, Turner:1997aa, Risaliti:1999aa, Garcet:2007aa, Mainieri:2007aa, Tajer:2007aa, Antonucci:2012aa, Malizia:2012aa, Merloni:2014aa, Davies:2015uq} in accordance with the unified model. Of course, the unified model, while broadly successful in explaining the diversity of AGN, is simplified and investigations of differences between Type 1 and Type 2 AGN that can not be explained by this paradigm can help to reveal the complex nature of AGN.

Of particular interest are the frequency and specific cases where the optical and X-ray classification disagree. Type 2 AGN that are X-ray unabsorbed have long been targets of study, and the debate over whether they represent AGN lacking a BLR is still ongoing \citep{Panessa:2002aa, Page:2006aa, Stern:2012fk, Merloni:2014aa}. Here, we focus on the opposite case, Type 1 AGN that appear to be X-ray absorbed.

Previous studies have found Type 1 AGN with large X-ray absorbing columns, however both the fraction and interpretation have varied. \citet{Perola:2004pd} found that 10\% of broad line AGN are X-ray absorbed within the HELLAS2XMM 1 degree field survey while \citet{Tozzi:2006sj} estimated at least 20\% of AGN in the \textit{Chandra} Deep Field South have inconsistent optical and X-ray classifications. Both \citet{Tajer:2007aa} and more recently \citet{Merloni:2014aa} instead find around 30\% of optically unobscured AGN are X-ray absorbed. \citet{Merloni:2014aa}, interestingly, also showed an increasing fraction of X-ray absorbed, but optically unobscured AGN at higher X-ray luminosities.

There are also several explanations for observing X-ray absorbed broad line AGN that only require small or no modifications to the unified model. An easy explanation is that our line of sight is grazing the edge of the torus where perhaps the cloud distribution is less dense but the covering fraction of the X-ray corona is much larger than the BLR due to the corona's smaller physical size. Another related possibility is that a cloud, perhaps from the torus or the BLR itself has entered our line sight causing a relatively brief increase in the X-ray absorbing column but leaving the BLR emission unaffected. Both explanations would also explain the relative rarity of X-ray absorbed Type 1 AGN. 

\citet{Davies:2015uq} suggested a luminosity dependence on the gas properties of the torus. At low luminosities, dust in the torus extends all the way down to the inner edge while at higher luminosities, dust-free gas dominates at small radii and changes to the standard dusty torus at larger radii. Thus, the X-ray absorbed fraction, does not change with luminosity but the optically obscured fraction should decrease with increasing luminosity as more lines of sight open up towards the BLR. This explains the luminosity dependencies seen in \citet{Merloni:2014aa} but keeps the popular unified model intact with only a slight modification.

What has been lacking in previous studies on the relationship between optical obscuration and X-ray absorption is relatively bias free selection of AGN and consistent classifications and measures of the AGN properties. Many of the studies previously mentioned have relied on AGN selection in the 2--10 keV band which can be heavily affected by even moderate X-ray absorption \citep[\nh $\sim10^{23}$][]{Koss:2016eh} and thereby biasing the sample against X-ray absorbed objects. Further, definitions of optically obscured and unobscured, as well as X-ray absorbed and unabsorbed, have either changed or relied on less reliable methods such as SED fitting to define optical obscuration, or hardness ratios to define the level of X-ray absorption.

In this work, we draw on a sample of low-redshift AGN selected at ultra-hard X-rays (14--195 keV) that largely avoid biases due to both X-ray absorption and host galaxy contamination. Further, this large sample of AGN has been systematically analyzed in both the optical and X-ray regime \citep{Koss:2017aa, Ricci:2017pl} leading to well-defined classifications and measurements of their properties, including the X-ray absorbing column and broad line emission. We use these measurements to investigate the prevalence of strong X-ray absorption in Type 1 AGN and discuss the implications of our results in the context of the unified model and the structure of AGN.

\section{Sample and Data}\label{sec:sample}

Our parent sample consists of all AGN in the BAT AGN Spectroscopic Survey\footnote{\url{https://www.bass-survey.com}} \citep[BASS;][]{Koss:2017aa, Ricci:2017pl}. The BASS team has analyzed both new and archival optical spectra for a large fraction (77\%; 641/836) of the AGN detected as part of the 70-month \textit{Swift} Burst Alert Telescope \citep[BAT;][]{Gehrels:2004qf, Barthelmy:2005ul} catalogue \citep{Baumgartner:2013fq}. \textit{Swift}/BAT has been continuously surveying the entire sky at high energies (14--195 keV) that produces nearly complete samples of AGN up to the Compton thick limit \citep{Ricci:2015aa} and reduces selection effects associated with host galaxy contamination and obscuration. \citet{Koss:2017aa} found that the average reddening, measured using the Balmer decrement, for the BASS AGN is significantly higher than optically selected AGN from SDSS and that a significant fraction of BASS AGN lacked any Balmer lines. This all points to hard X-ray selected AGN samples including more obscured or optically contaminated AGN that optical spectroscopic surveys would not select.

For this work, we need reliable measurements of the broad \halpha{} flux, intrinsic hard X-ray flux, and X-ray absorbing column density. Therefore, we chose all AGN with detected broad \halpha{} from the original BASS analysis as well as AGN that were part of the BASS X-ray spectral analysis presented in the BASS X-ray catalog \citep{Ricci:2017pl}. The key measurements obtained from the X-ray spectral analysis are \nh{} estimates and k- and absorption corrected 14--150 keV flux (hereafter referred to as the intrinsic X-ray flux). Details of the X-ray spectral analysis can be found in \citet{Ricci:2017pl}.

We limited our sample based on the following requirements with fractions of the parent sample included in parentheses:
\begin{enumerate}\itemsep1pt
\item Seyfert classification according to \citet{Winkler:1992kx} (594/641, 93\%, see below and Appendix~\ref{app:outflows} for our modifications)
\item Non-blazar based on exclusion from the Roma Blazar Catalog \citep{Massaro:2009aa} (581/641, 91\%)
\item X-ray flux and \nh{} measurement (638/641, 99\%)
\item Measured distance (634/641, 99\%)
\item Quality flag of 1 or 2 for the spectral fitting of the \halpha{} region (226/641, 35\%)
\end{enumerate}

The last requirement above regarding quality flags ensured we removed all sources whose spectra either did not cover the \halpha{} spectral region or the spectral fitting were unreliable. For a detailed description of the optical spectral fitting and explanation of each flag, see the BASS Data Release 1 publication \citep{Koss:2017aa}. Briefly, a quality flag of 1 indicates a good fit with small residuals while a quality flag of 2 indicates an acceptable fit with larger residuals but overall representative of the emission line profiles. As expected the quality flag requirement is the most restrictive given it filters out nearly all Type 2 AGN as well as some Type 1 AGN with either poor spectra or high redshifts that move \halpha{} out of the spectral range. 

Broad \halpha{} and intrinsic X-ray fluxes were converted to luminosities using either the redshift independent distances compiled from the NASA/IPAC Extragalactic Database\footnote{\url{http://ned.ipac.caltech.edu/}} when available or luminosity distances calculated based on the measured redshifts from the spectral analysis and our chosen cosmology ($H_{0}=70$ km s$^{-1}$ Mpc$^{-1}$, $\Omega_{m} = 0.3$).

Finally, we wanted to ensure the measured broad \halpha{} component was truly originating from the BLR. \citet{Trippe:2010aa} showed that intermediate type AGN, especially Sy 1.8s and 1.9s, can often be misclassified. While Sy 1--1.5s have a corresponding broad H$\beta$ to match their broad \halpha{} component, Sy 1.9s by their very definition do not. Therefore, it is possible that what may seem like a broad \halpha{} component could in fact be due to another process entirely unrelated to the BLR such as an outflow especially for sources with low measured FWHM ($\sigma_{\rm v} < 2000$ km s$^{-1}$) for broad \halpha{}. High velocity wings present in both \halpha{} and [NII] could be disguised as a faint broad \halpha{} component in moderate resolution spectra due to the heavy blending between \halpha{} and the [NII] doublet. In Appendix~\ref{app:outflows}, we reanalyzed the \halpha+[NII] complex for the  57 Sy 1.9s that originally fit the above criteris and determined that for 18/57 ($\sim30$\%) Sy 1.9s, an outflowing component could reasonably explain the originally measured broad \halpha{} component. While not definitive proof that these AGN have been misclassified, they could potentially bias our results, especially since many are X-ray absorbed. Thus, we have chosen to remove these 18 Sy 1.9s from our sample.

Our final sample consists of 190 Type 1 AGN with 20, 66, 65, and 38 Sy 1s, 1.2s, 1.5s, and 1.9s respectively\footnote{The BASS sample interestingly does not contain any Sy 1.8s.}.

\section{Results}\label{sec:results}
\begin{figure}
\includegraphics[width=\columnwidth]{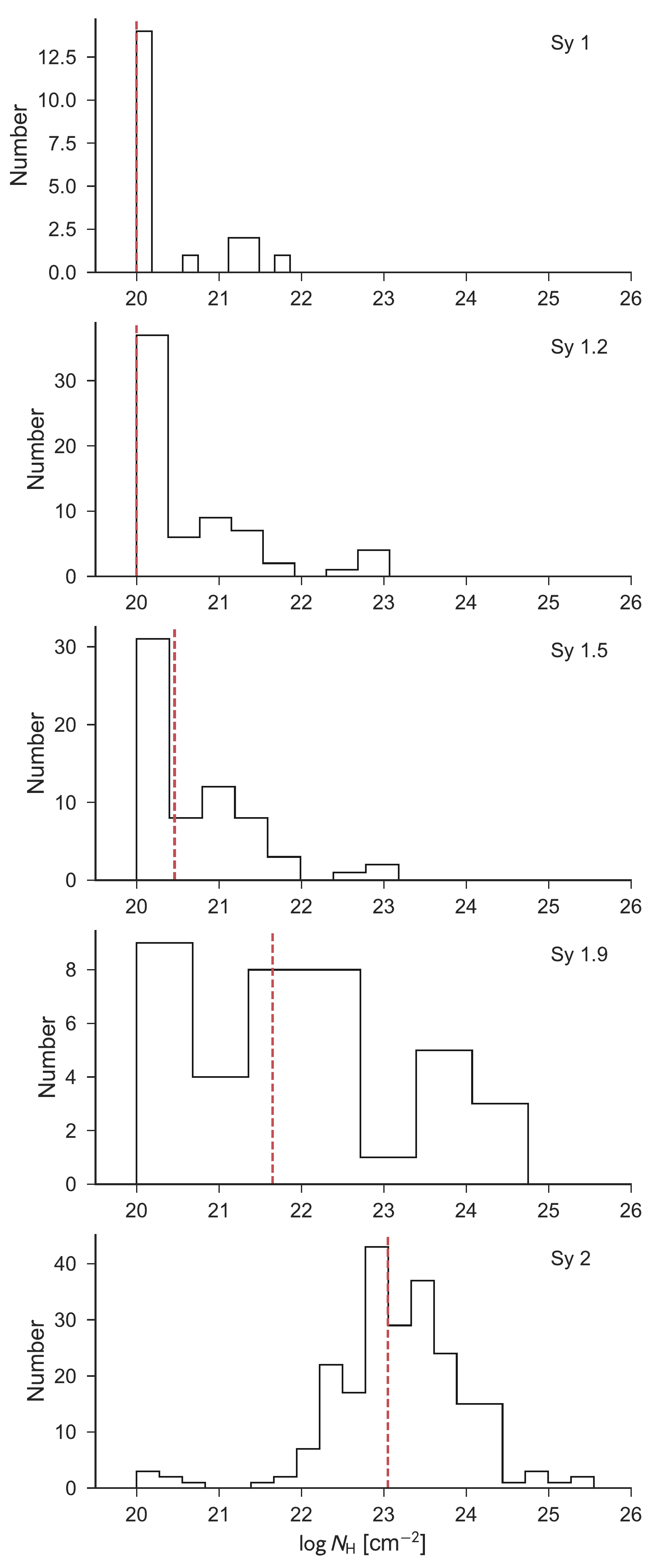}
\caption{\label{fig:nh_dist}\nh{} distribution for the Sy 1s, 1.2s, 1.5s, 1.9s in our Type 1 AGN sample and Sy 2s in the full BASS sample. Red dashed lines indicate the median \nh{} for each subsample excluding \nh$>10^{24}$~cm$^{-2}$ to avoid incompleteness. \nh$=10^{20}$~cm$^{-2}$ is the lowest column density that is able to be measured in the X-ray due to Galactic absorption.}
\end{figure}

We first start by showing the \nh{} distribution for the our Type 1 AGN sample in Fig.~\ref{fig:nh_dist} regardless of whether the source has a broad \halpha{} measurement and include as well all of the Sy 2s from BASS for reference. While there is an increase in the number of X-ray absorbed AGN moving from Sy 1s to 2s, the biggest increase certainly occurs for the Sy 1.9 subsample. If we arbitrarily create a cutoff at \nh$=10^{22}$ cm$^{-2}$ for X-ray unabsorbed and absorbed AGN, the fraction of X-ray absorbed AGN is $<0.12$, 0.08$^{+0.08}_{-0.05}$, 0.05$^{+0.16}_{-0.03}$, 0.45$^{+0.16}_{-0.15}$, and 0.96$^{+0.03}_{-0.02}$ for Sy 1s, 1.2s, 1.5s, 1.9s, and 2s respectively where the uncertainties have been calculated assuming binomial statistics \citep{Gehrels:1986sf}. Therefore, the level of X-ray absorption in Sy 1.9s seems to be intermediate between Sy 1-1.5s and Sy 2s based on the fraction of sources with high X-ray absorption. This is certainly not new and has been observed before in smaller samples \citep[e.g.][]{Risaliti:1999aa} and is the reason Sy 1.9s are routinely grouped along with Sy 2s to form a general ``absorbed'' AGN sample.

Na\"{i}vely, one would expect the same structure (i.e. dusty torus) to cause both optical obscuration of the BLR and absorption of the X-ray emission. We would then expect to observe a level of optical obscuration in Sy 1.9s consistent with the X-ray \nh{} measurements. Measuring the optical obscuration towards the BLR however is difficult and usually involves a number of assumptions about the geometry and ionization state of BLR clouds. The standard method is to assume Case B recombination and use the ratio of broad H$\alpha$ to H$\beta$ emission as an estimate of the BLR extinction. By the very definition of a Sy 1.9 AGN (i.e.~absence of broad H$\beta$), however, this method is not viable for all sources of our sample because it can provide only lower limits to the optical obscuration. Further, the assumption of Case B recombination in the BLR has been shown to be a questionable assumption \citep[e.g.][]{Schnorr-Muller:2016qy}. 

Instead, we rely on the existence of a linear relationship between the bolometric AGN luminosity and the broad \halpha{} luminosity. \citet{Stern:2012fk} studied  more than 3000 broad line AGN from the \textit{Sloan Digital Sky Survey} (SDSS), finding that the relation between far-UV (near the peak of the AGN SED) and broad \halpha{} is linear especially for the highest luminosity bins. At lower broad \halpha{} luminosity, the relation does slightly flatten such that they observe more FUV emission than expected; however the deviation is small and they show that it is likely due to host galaxy contamination based on the changing broad \halpha{} relationships with other wavelengths. Thus, they conclude that the covering fraction of the BLR is likely independent of AGN luminosity and the broad \halpha{} luminosity can reliably be used as a tracer of the bolometric luminosity. Especially useful for this work is the fact that in their analysis the tightest correlation with broad \halpha{} occurred with the 2 keV monochromatic luminosity which is the most likely regime to be unaffected by any host galaxy contribution.

\citet{Elitzur:2014aa} dispute this and show that by subclassing the SDSS broad line AGN, intermediate-type AGN exhibit reduced levels of broad \halpha{} compared to their X-ray emission. They interpret this as a reduction in the covering factor ($\propto L_{\rm bH\alpha}/L_{\rm Bol}$) for the BLR as the Eddington ratio ($\propto L_{\rm Bol}/M_{\rm BH}$) decreases which they also find is similarly reduced for intermediate type AGN.  Our need, however, is only an upper limit on the optical obscuration towards the BLR. If the intrinsic relationship is linear then we should see a reduction in $L_{\rm bH\alpha}/L_{\rm Bol}$ and can use the deficit as a measure of the extinction. If, on the other hand, the intrinsic relationship is not linear and $L_{\rm bH\alpha}/L_{\rm Bol}$ decreases with luminosity, then the deviation from a linear relationship is not indicative of any optical obscuration and the extinction towards the BLR is negligible. Therefore, by assuming a linear relationship, we are actually being conservative in our measurement of the BLR extinction.

\begin{figure*}
\centering
\includegraphics[width=0.8\textwidth]{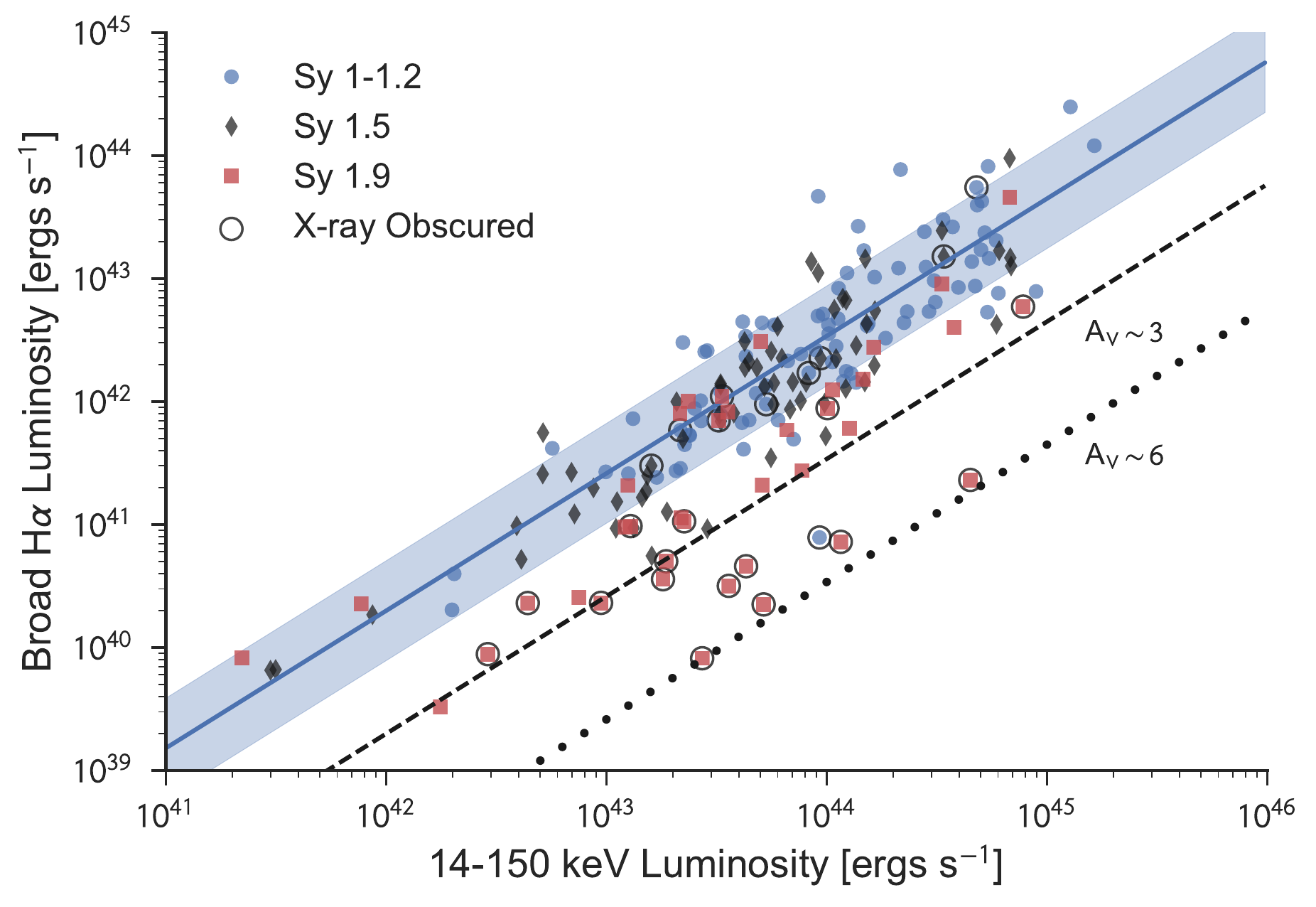}
\caption{\label{fig:brha_xray} The relationship between intrinsic X-ray (14--150 keV) luminosity and observed broad \halpha{} luminosity for BASS selected Sy 1--1.9s. Blue dots correspond to Sy 1--1.2s which were used to measure the best-fit line (solid blue line) and scatter (blue shaded region) between the X-ray emission and broad \halpha{} emission. Black diamonds show the Sy 1.5s while red squares plot the Sy 1.9s. X-ray absorbed sources (\nh{} $>10^{22}$ cm$^{-2}$) are encircled. The black dashed line shows our chosen threshold for optically obscured sources which are 1 dex (corresponding to about 2.5$\sigma$ and $A_{\rm V}\sim3$ mag) below the measured best-fit line while the black dotted line shows approximately the maximum reduction we observe in broad \halpha{} of 2 dex and $A_{\rm V}\sim6$ mag.}
\end{figure*}

To begin, we assume the intrinsic X-ray luminosity is an accurate tracer of the bolometric AGN luminosity as has been shown in previous studies \citep[e.g.][]{Vasudevan:2007kx, Winter:2012yq}. In Fig.~\ref{fig:brha_xray}, we show the correlation between the broad \halpha{} and intrinsic X-ray luminosities for Sy 1-1.2s as blue dots. The correlation is highly significant with a Pearson correlation coefficient of 0.85 and p-value of $<0.001$. We fit a simple line using linear least squares between the broad \halpha{} luminosity and intrinsic X-ray luminosity finding the following relation:

\begin{equation}\label{eq:brha-xray}
\log\,L_{\rm bH\alpha} = 1.06\log\,L_{\rm 14-150\,keV} - 4.32\footnote{Using a photon index of 1.8, the equivalent relationship for the 2--10 keV energy range has an intercept of 4.71}
\end{equation}

We plot Equation~\ref{eq:brha-xray} in Fig.~\ref{fig:brha_xray} as a blue line along with shading to indicate our measured $\pm0.4$ dex scatter. As red squares we show our Sy 1.9 sample. A large fraction of Sy 1.9s lie systematically below the relationship defined by the Sy 1-1.2s and well outside the estimated scatter. The black dotted line indicates a reduction of 2 dex in the broad \halpha{} luminosity and highlights roughly the maximum decrease we observe for Sy 1.9s. We also plot Sy 1.5s as black diamonds to show that while they were not included in our calculation of the best fit, Sy 1.5s also lie along the line and within the scatter. Finally, sources outlined by black circles are X-ray absorbed, defined above as having \nh{} $>10^{22}$ cm$^{-2}$.

Assuming that the reduction in broad \halpha{} luminosity is completely due to obscuration, we calculate the visual extinction, $A_{\rm V}$, given an extinction law. For this work, we use the empirically determined extinction law from \citet{Wild:2011aa} that was also used in our previous study investigating BLR obscuration \citep{Schnorr-Muller:2016qy}:

\begin{equation}\label{eq:wild_extinct_law}
\frac{A_{\lambda}}{A_{\rm V}} = 0.6(\lambda/5500)^{-1.3} + 0.4(\lambda/5500)^{-0.7},
\end{equation}
\noindent where $\lambda$ is the rest wavelength for \halpha{} (6563 \AA) and  $A_{\rm bH\alpha} = -2.5\log(L_{\rm bH\alpha, X}/L_{\rm bH\alpha. obs})$ . In the last relation $L_{\rm bH\alpha, X}$ is the expected broad \halpha{} luminosity based on Equation~\ref{eq:brha-xray} and $L_{\rm bH\alpha, obs}$ is the observed broad \halpha{} luminosity. Using these equations, the maximum reduction in the broad \halpha{} luminosity we observe (black dotted line) corresponds to only an $A_{\rm V} \sim 6$ mag. If we also assume a Galactic \nh/$A_{\rm V}$ ratio of $1.87\times10^{21}$ cm$^{-2}$ \citep{Draine:2011zr}, we expect a maximum \nh{}$\sim10^{22}$ cm$^{-2}$, the cutoff we used for our definition of X-ray absorbed. Therefore, based on the optical obscuration towards the BLR, we would expect virtually all of our Sy 1.9s to be X-ray unabsorbed. Instead, Figs.~\ref{fig:nh_dist} and \ref{fig:brha_xray} show that over half of the Sy 1.9s have \nh{} values above 10$^{22}$ cm$^{-2}$ meaning along our lines of sight towards a significant fraction of Sy 1.9s, there is either more or different gas and dust hiding the central X-ray source than there is in front of the BLR.

\begin{figure*}
\centering
\includegraphics[width=0.8\textwidth]{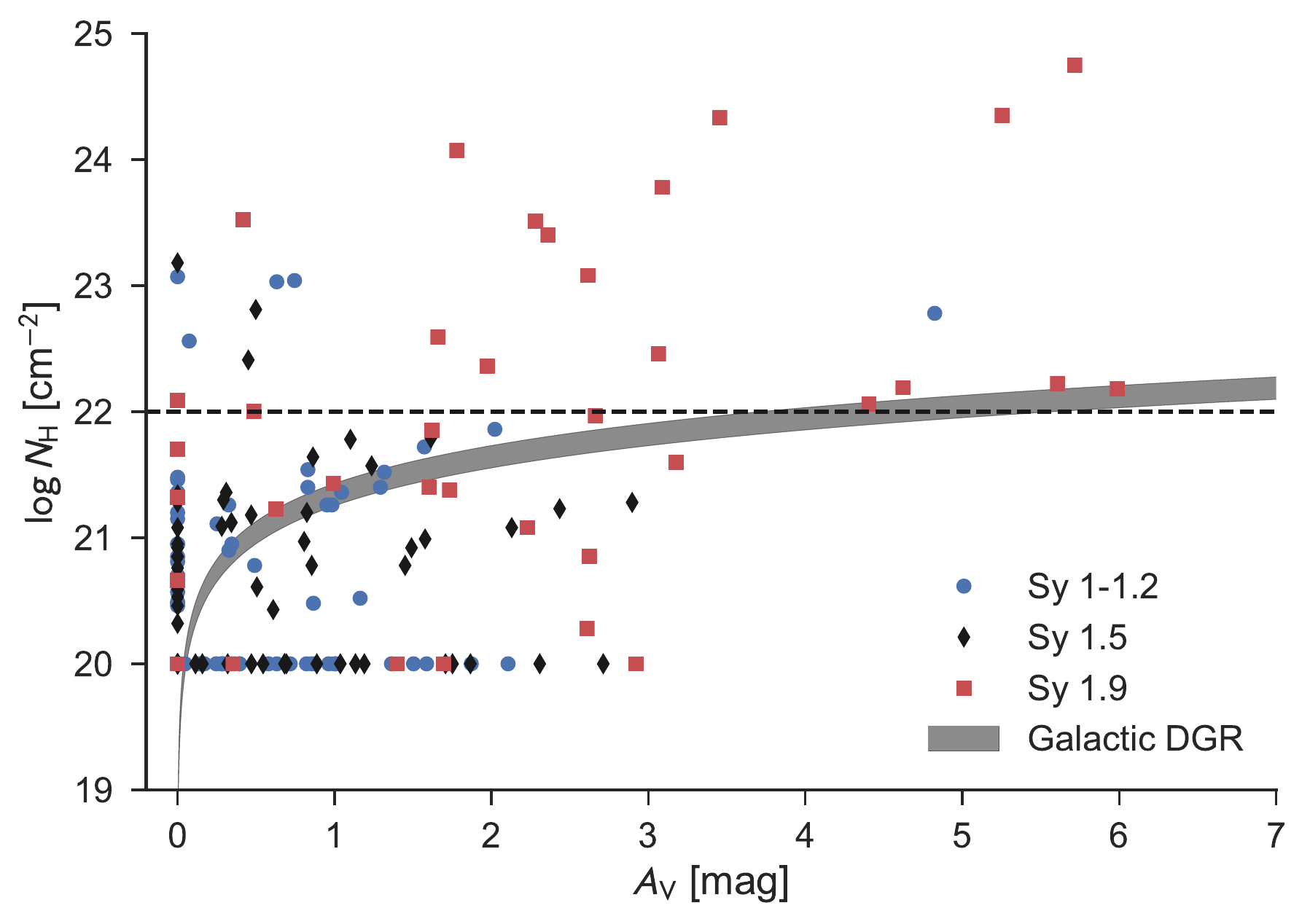}
\caption{\label{fig:nh_vs_av}A comparison between the X-ray absorbing column density (\nh) and the optical extinction measured using the broad \halpha-to-X-ray relationship. The symbols are the same as in Fig.~\ref{fig:brha_xray} and the gray shaded region is the expected relationship for the range of Galactic dust-to-gas ratios found in the literature (\nh/$A_{\rm V} = 1.79 - 2.69 \times 10^{21}\,\rm{cm}^{-2}$). The black dashed line indicates the usual cutoff discriminating X-ray unabsorbed and absorbed AGN.}
\end{figure*}

This is also shown by comparing X-ray derived \nh{} values with the broad \halpha{} derived $A_{\rm V}$ values illustrated in Fig.~\ref{fig:nh_vs_av}. For most of the Sy 1--1.5s, both \nh{} and $A_{\rm V}$ are low whereas the Sy 1.9s form the majority of the high \nh{} and high $A_{\rm V}$ Type 1 AGN. Below $A_{\rm V} = 3$ mag, most of the AGN either scatter around the gray shaded line, which represents the spread of typical gas-to-dust ratios (DGR) found in our Galaxy (\nh/$A_{\rm V} = 1.79 - 2.69 \times 10^{21}\,\rm{cm}^{-2}$), or lie along \nh$=10^{20}\,\rm{cm}^{-2}$ or $A_{\rm V} = 0$ mag. Above $A_{\rm V} = 3$ mag, all but one AGN is a Sy 1.9 and all either lie on the Galactic DGR line or above it, sometimes with several orders of magnitude more column density than expected for a Galactic DGR. 

Fig.~\ref{fig:nh_vs_av} confirms the findings of \citet{Schnorr-Muller:2016qy} and \citet{Burtscher:2016fk}. Both of these studies, through independent methods, found that intermediate Seyferts typically display moderate optical obscuration ($A_{\rm V} = 4-8\,\rm{mag}$). The similarity between the two previous studies and ours validates our relatively simple method for measuring the optical obscuration towards the BLR. With our much larger sample, though, we find a much larger range in \nh{} values for Sy 1.9s. Whereas \citet{Burtscher:2016fk} determined that using $\log$\nh$=22.3\,\rm{cm}^{-2}$ as a threshold for X-ray absorbed AGN consistently classified Sy 1.9s as unobscured objects, we find that Sy 1.9s instead span $\log$\nh{} values all the way up to 25. Indeed, 12/39 (32\%) of Sy 1.9 have \nh{} measurements above this threshold. Sy 1.9s represent 60\% of the Type 1 AGN that can be considered X-ray absorbed using a threshold of $\log$\nh$>22.3\,\rm{cm}^{-2}$, which leads to a total X-ray absorbed, Type 1 AGN frequency of 10\%. Lowering the threshold for X-ray absorbed AGN to $\log$\nh$=21.5\,\rm{cm}^{-2}$ increases the frequency to 20\% so we can confidently say the X-ray absorbed fraction within Type 1 AGN is between 10 and 20\%. These fractions are more in agreement with the results of \citet{Perola:2004pd} and nearly a factor three smaller than the rates found by \citet{Tajer:2007aa} and \citet{Merloni:2014aa}. What is further clear from this study is that the hydrogen column densities measured from the X-ray spectra are generally much larger than those measured from the BLR extinction assuming a Galactic GDR. 

\section{Discussion}\label{sec:discuss}

\subsection{Comparison with \halpha/H$\beta$ ratios}
An important question in our analysis is whether our estimates of the optical extinction are reliable. While the values of $A_{\rm V}$ seem to match those found in previous studies using independent methods, we can still test whether our measurements of $A_{\rm V}$ make sense based on the Balmer decrement. Our first test assumes that the intrinsic \halpha/H$\beta$ ratio in the BLR is 3.1, i.e. the ratio for Case B recombination, adjusted for the recombination of helium. 

For the Sy 1--1.5s, we can simply use the measured broad H$\beta$ from BASS DR1. For the Sy 1.9s, where broad H$\beta$ is absent, we derived upper limits using a Monte Carlo Markov Chain (MCMC) analysis on the H$\beta$ spectral region with the Python package \textsc{EMCEE} \citep{Foreman-Mackey:2013lr}. We fixed the FWHM and velocity of the narrow H$\beta$ component to the values from BASS DR1 and fixed the FWHM for the broad H$\beta$ component to that of broad \halpha. Therefore, the only free parameters in the modeling is the amplitude of the broad and narrow component and the velocity of the broad component. For these parameters, we simply used a uniform prior between 0 and infinity for the amplitudes, and allowed the line center to vary between -1000 and 1000 km/s. Upper limits on the broad H$\beta$ flux were then calculated using the 99th percentile of the marginalized probability distribution for the broad H$\beta$ amplitude, which equates to about a 3$\sigma$ upper limit. 

\begin{figure}
\centering
\includegraphics[width=\columnwidth]{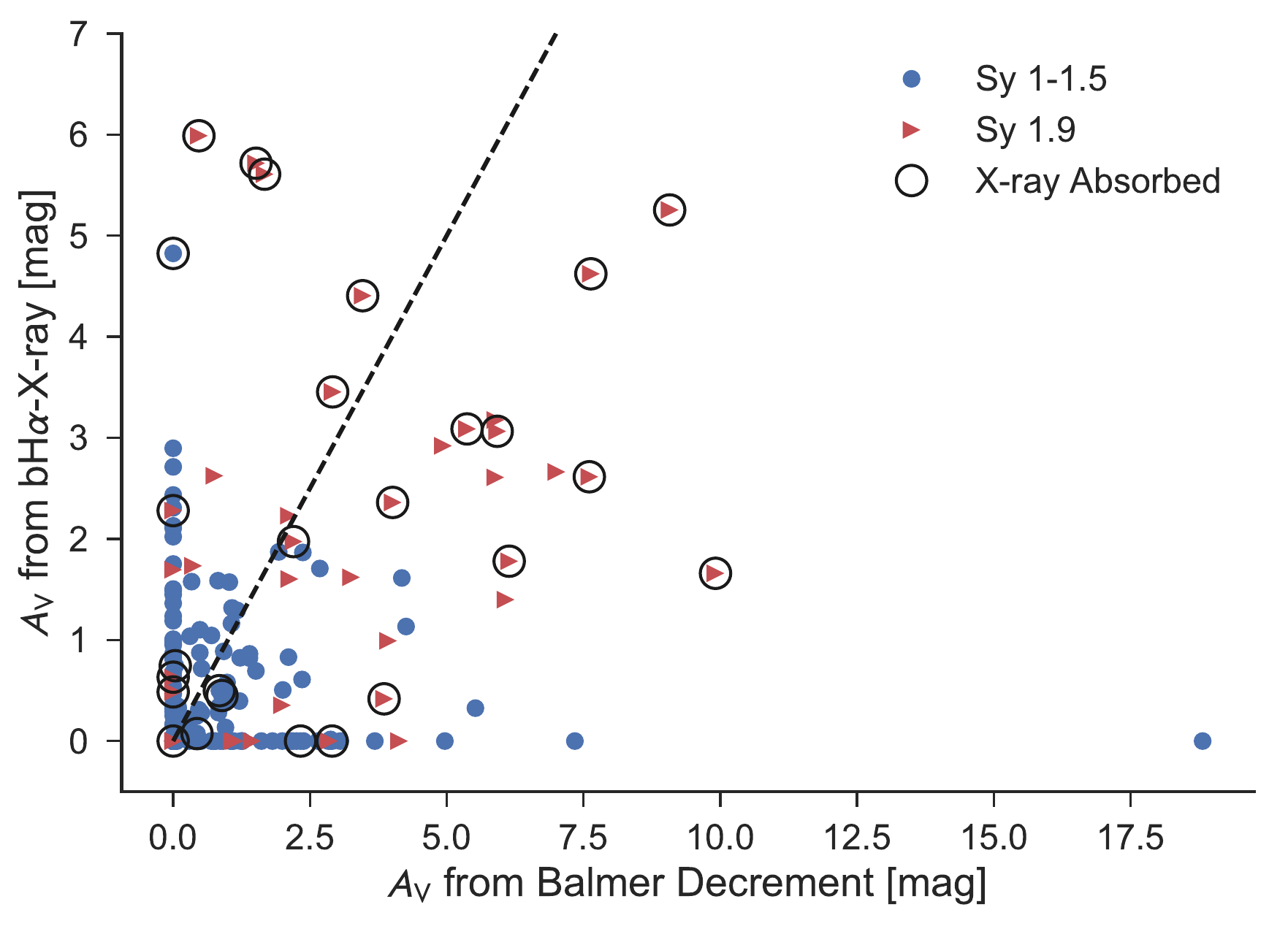}
\caption{\label{fig:compare_av} Comparison between the optical extinction measured from the broad \halpha-to-X-ray relationship and the Balmer decrement. Blue points indicate Sy 1--1.5s for which measurements of broad H$\beta$ were made while the right-facing red triangles indicate Sy 1.9s where only a 3$\sigma$ upper limit for broad H$\beta$ could be obtained so the $A_{\rm V}$s from the Balmer decrement are lower limits. The dashed line indicates a 1-1 relationship.}
\end{figure}

$A_{\rm V}$ from the Balmer decrement were calculated using the same extinction curve given in Equation~\ref{eq:wild_extinct_law}. Figure~\ref{fig:compare_av} shows the comparison between the $A_{\rm V}$ determined from the broad \halpha-to-X-ray relationship and the Balmer decrement. Sy 1--1.5s are shown as blue dots while the Sy 1.9s are shown as right-pointing triangles to signify the $A_{\rm V}$ from the Balmer decrement are lower limits. Black circles show X-ray absorbed AGN and the black dashed line indicates a 1-1 correspondence.

\begin{figure*}
\centering
\includegraphics[width=\textwidth]{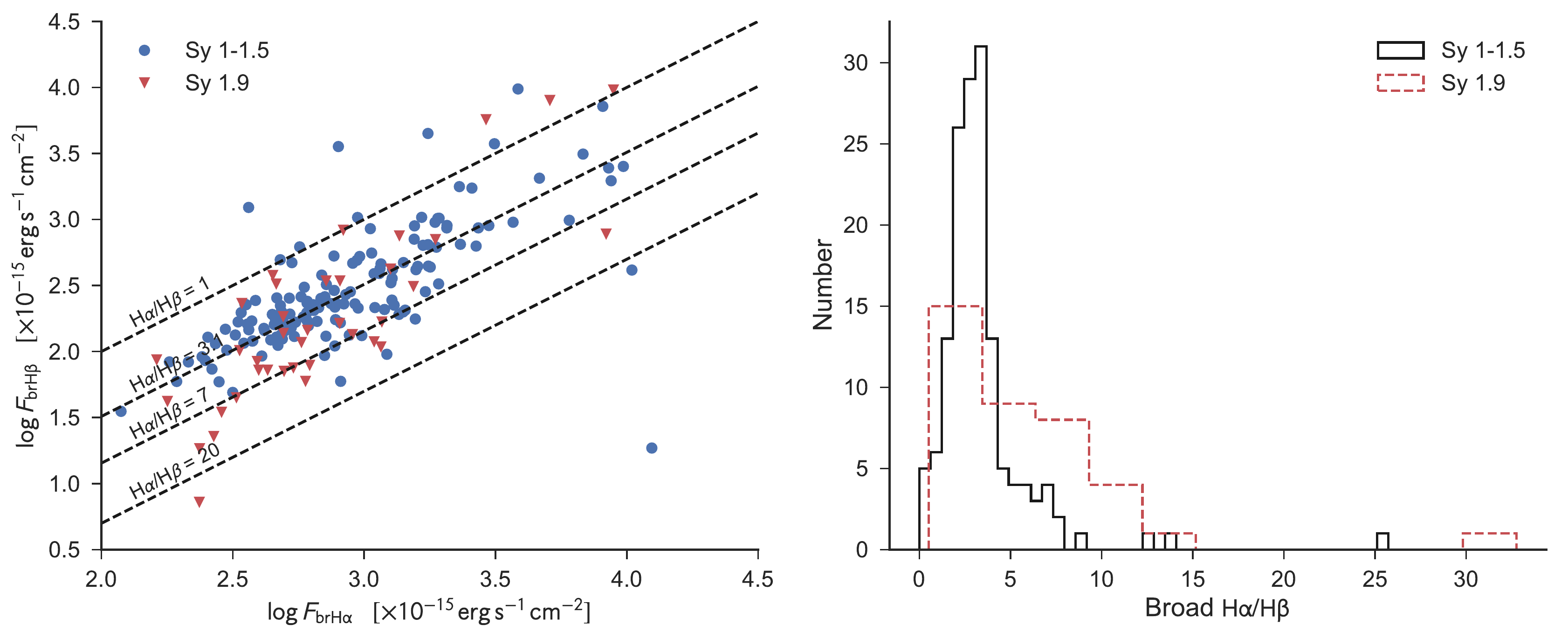}
\caption{\label{fig:halpha_hbeta_ratios}\textit{Left:} Correlation between broad \halpha{} and broad H$\beta$ for Sy 1--1.5s (blue points) and Sy 1.9s (red triangles) after correcting for extinction under the assumption the values of $A_{\rm V}$ from the broad \halpha-to-X-ray relationship are correct. As in Fig.~\ref{fig:compare_av}, the broad H$\beta$ measurements for Sy 1.9s are 3$\sigma$ upper limits. The black dotted lines show lines of constant \halpha/H$\beta$ ratio for various ratios between 1 and 20. \textit{Right:} Distribution of intrinsic broad \halpha/H$\beta$ ratios for Sy 1--1.5s (black solid line) and lower limits for Sy 1.9s (red dashed line). }
\end{figure*}

Sy 1--1.5s seem to show broad agreement between the two methods. Nearly all of them are between 0--3 mag for both $A_{\rm V}$ measurements. Sy 1.9s on the other hand are well scattered across the 1-1 correspondence line with a small indication that the Balmer decrement method is producing higher $A_{\rm V}$ compared to the broad \halpha-to-X-ray relationship, especially considering all of these values are corrected lower limits. 

However, the previous analysis assumes an intrinsic ratio of 3.1, while multiple studies have indicated a large variation in the intrinsic \halpha/H$\beta$ ratio for AGN \citep[e.g.][]{Baron:2016aa, Schnorr-Muller:2016qy}. Therefore, for this next test, we instead assume that our measured optical extinction from the broad \halpha-to-X-ray relationship is correct, and use it to infer the intrinsic line ratio (or lower limit for Sy 1.9s) for our object. Figure~\ref{fig:halpha_hbeta_ratios} shows the results with the left panel displaying the relationship between the broad \halpha{} and H$\beta$ flux (upper limits for Sy 1.9s) for individual objects. Dashed lines indicate lines of constant line ratio. The right panel shows the distribution of inferred intrinsic \halpha/H$\beta$ ratio for Sy 1--1.5s (black line) and lower limits for Sy 1.9s (red dashed line). We find that for our Type 1 AGN, our derived optical extinctions result in intrinsic ratios of 1--7 with most objects lying along the Case B value of 3.1. This range of \halpha/H$\beta$ ratios is consistent with the range seen by \citet{Schnorr-Muller:2016qy}. Lower limits for Sy 1.9s do show a stronger tail towards larger ratios with a higher fraction of sources reaching a ratio of $\sim$10, however still largely distributed around a value of 3.1. We conclude from this that our estimates of $A_{\rm V}$ result in intrinsic line ratios consistent with those observed in previous studies.

\subsection{Implications on the structure and geometry of AGN}

\begin{figure}
\includegraphics[width=\columnwidth]{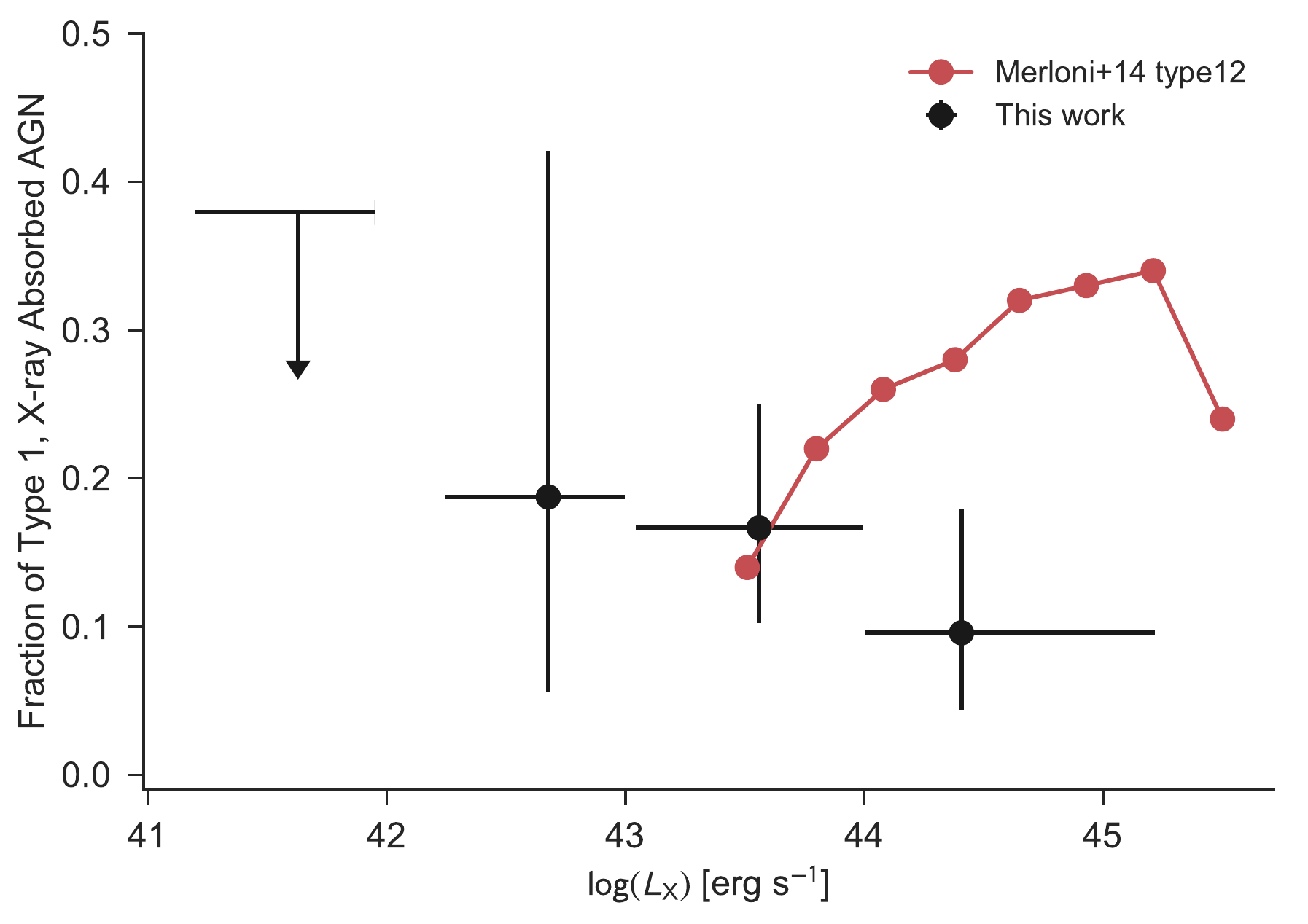}
\caption{\label{fig:frac_xabs} The fraction of X-ray absorbed Type 1 AGN as a function of intrinsic X-ray luminosity for the BASS sample. Error bars and upper limits indicate the 95\% binomial confidence interval. The red points and line show the same fraction found in the higher redshift sample from \citet{Merloni:2014aa}. While the fraction of X-ray absorbed Type 1 AGN increases strongly with luminosity in the high redshift sample, our low redshift sample fraction either remains constant or perhaps decreases.}
\end{figure}

One possible implication from our simple analysis is that the dust and gas obscuring the central X-ray corona in Type 1 AGN is internal to the BLR. In fact, as suggested in several studies \citep{Merloni:2014aa, Davies:2015uq, Burtscher:2016fk} , the X-ray obscuring structure is the neutral, dust-free gas within the BLR itself, a so-called ``neutral torus'' that is the inner extension of the dusty and molecular torus that creates the optical obscuration.  These X-ray absorbed Type 1 AGN, then are seen along lines of sight through the neutral torus, but not the dusty molecular torus which would lead to a standard Type 2 AGN. 

A key prediction for this scheme is an increase in the number of X-ray absorbed, Type 1 AGN as a function of AGN luminosity. This would occur due to the increase of the dust sublimation radius. At low luminosity, the dust sublimation radius is closer in, reducing the fraction of lines of sight that only intersect the neutral torus but not the molecular torus. We can check this prediction with our Type 1 sample. We determined the fraction of X-ray absorbed Type 1 AGN within four $\log\,L_{\rm X}$ bins and show the results in Fig.~\ref{fig:frac_xabs}. The error bars on the fraction represent the 95\% binomial confidence interval while the error bars on the X-ray luminosity represent the range within the bin. In the lowest luminosity bin between $\log\,L_{\rm X}=41-42$ erg s$^{-1}$ there are no X-ray absorbed Type 1 AGN, but there are also only five total Type 1 AGN as we are hampered by the flux limit of the BAT survey. Therefore, we show the 95\% confidence upper limit. In the second bin, 4/16 Type 1 AGN are X-ray absorbed and the large error bars reflect the relatively small sample size. 

Focusing on the two largest luminosity bins which, centered on $\log\,L_{\rm X} = 43.5$ and 44.4 and containing 96 and 73 AGN respectively, we do not find a clear increase in the X-ray absorbed fraction as expected if the extent of the neutral torus is increasing with higher luminosity. Instead, it appears the fraction is constant or possibly decreases. Using Fisher's exact test, we find a p-value of 0.36 indicating we cannot reject the null hypothesis that the X-ray absorbed fraction is the same in both bins. In fact, the p-value increases to 0.87 under the null hypothesis that the X-ray absorbed fraction is \textit{less} than the lower luminosity bin. The results do not change if we combine all three of the lowest luminosity bins into one single bin and compare it to the highest luminosity bin.

This is also evident in Fig.~\ref{fig:frac_xabs} which also shows a comparison between the results obtained from the BASS sample and that from \citet{Merloni:2014aa}. The ``type 12'' AGN from their sample are the same X-ray absorbed Type 1 AGN studied here except at higher redshifts ($0.3<z<3.$). There is a clear rise in the type 12 fraction as a function of X-ray luminosity that is not reflected in our low redshift sample. Several factors could account for the discrepancy. The parent BAT AGN sample from which our study is based only covers a relatively small volume compared to the \citet{Merloni:2014aa} sample. Therefore, our study does not include many high luminosity AGN since the number density drops rapidly although our last bin contains 73 AGN. Since high quality X-ray spectra were not available for all sources, \citet{Merloni:2014aa} relied on hardness ratios to determine the X-ray absorbing column density. As they show, this method has a large scatter when compared to spectral measurements with differences up to 2 dex possible. As such, it is currently unclear whether the X-ray absorbed, Type 1 fraction increases, decreases, or is constant at higher luminosities. 

If we suppose our measurements, completely determined from both optical and X-ray spectra, are correct, then we can put them into the context of the recent work of \citet{Ricci:2017ek}. They studied the general X-ray obscured fraction in the entire BASS sample, finding a significant decrease of the total X-ray absorbed fraction at high Eddington ratio ($\lambda_{\rm Edd}$). The explanation in \citet{Ricci:2017ek}, is that radiative feedback from the AGN shapes the obscuring structure. At low Eddington ratio, gas and dust are able to build up around the SMBH, increasing the covering factor of the ``torus'', while at high Eddington, the AGN has cleared away large amounts of gas and dust. This results in a dramatic increase in unobscured AGN at high Eddington as more lines of sight towards the BLR open up. 

For Type 1 AGN, we do not observe a similar radical decrease in the X-ray absorbed fraction that \citet{Ricci:2017ek} find for primarily Type 2 AGN. Using SMBH masses from the BASS DR1 and a cutoff of $\log\,\lambda_{\rm Edd} = -1.5$ , we find an X-ray absorbed fraction of 28$^{+16}_{-13}$\% and 18$^{+7}_{-5}$\% for low and high Eddington Type 1 AGN. This indicates that while the opening angle of the torus increases at high Eddington ratios, the fraction of sight lines \textit{through} the BLR only mildly decreases and suggests the covering factor of the BLR remains relatively constant. This could be further proof that dust is the key component to couple the AGN's power to its surrounding environment as the dust covering factor seems to respond more dramatically than the neutral, dust-free BLR. 

\section{Summary and Conclusions}\label{sec:conclude}
In this paper, we have examined the X-ray absorbed fraction of Type 1 AGN within a large, hard X-ray selected sample of low redshift AGN. Using the relationship between the broad \halpha{} and X-ray luminosity as an estimate of the optical extinction, we show the column densities of gas towards the X-ray corona and BLR are largely discrepant, indicating the X-ray absorbing material is either internal or coincident with the BLR. The following summarizes our results:

\begin{itemize}
\item Over the whole BASS sample, the fraction of Type 1 AGN (i.e. those that show at least broad \halpha), that are X-ray absorbed is between 10--20\% depending on the chosen \nh{} cutoff.
\item Up to 30\% of Sy 1.9s could be misclassified due to high velocity outflows masquerading as a BLR component.
\item The X-ray absorbed Type 1 fraction is relatively constant indicating a constant BLR covering fraction.
\item This further leads to a slight decrease with Eddington ratio, similar but not as dramatic as what is seen for the total fraction of obscured AGN in the entire BASS sample. This could be an indication that dust is a necessary ingredient for coupling AGN radiation to the surrounding ISM.
\end{itemize}

\acknowledgments
CR acknowledges financial support from the CONICYTChile grants FONDECYT 1141218 and Basal-CATA PFB--06/2007, and from the China-CONICYT fund. K.O. and K. S. acknowledge support from the Swiss National Science Foundation (SNSF) through Project grants 200021\textunderscore157021. M.K. acknowledges support from the SNSF through the Ambizione fellowship grant PZ00P2\textunderscore154799/1 and SNSF grant PP00P2 138979/1. KS acknowledges support from Swiss National Science Foundation Grants PP00P2\_138979 and PP00P2\_166159.

\software{%
	\textsl{astropy} \citep{Astropy:2013ek},
	\textsl{pandas} \citep{McKinney:2010em},
	\textsl{matplotlib} \citep{Hunter:2007},
	\textsl{numpy} \citep{vanderWalt:2011we},
	\textsl{scipy} \citep{Jones:2001ch}
}

\appendix
\section{Reevaluating the Broad \halpha{} Component for Sy 1.9}\label{app:outflows}
We investigate the possibility that some Sy 1.9s, especially those with small FWHMs for their broad \halpha{} component, could instead be Sy 2s with a strong outflowing component. The high-velocity wings associated with the outflow would be present in both the \halpha{} and [NII] line profiles and could be misinterpreted as an underlying broad \halpha{} component associated with the BLR. This could partly explain the high \nh{} values seen for a large fraction of Sy 1.9s.

NGC 5728 is a prime example of this misclassification. Within BASS, NGC 5728 was found to be a Sy 1.9 with a FWHM in broad \halpha{} of 1766 km s$^{-1}$. X-ray spectral analysis finds $\log\,N_{\rm H} = 24.13$, a seemingly perfect case of an AGN whose optical obscuration is much lower that the X-ray absorption. NGC 5728 has also been observed with VLT/X-Shooter and VLT/SINFONI as part of our ongoing Local Luminous AGN with Matched Analogues \citep[LLAMA][]{Davies:2015uq} program providing UV-NIR spectra with high spectral resolution ($R\sim8000$) and NIR H+K band integral field unit imaging with high spatial resolution ($\sim 0.15\arcsec$). 

\begin{figure}
\centering
\includegraphics[width=\columnwidth]{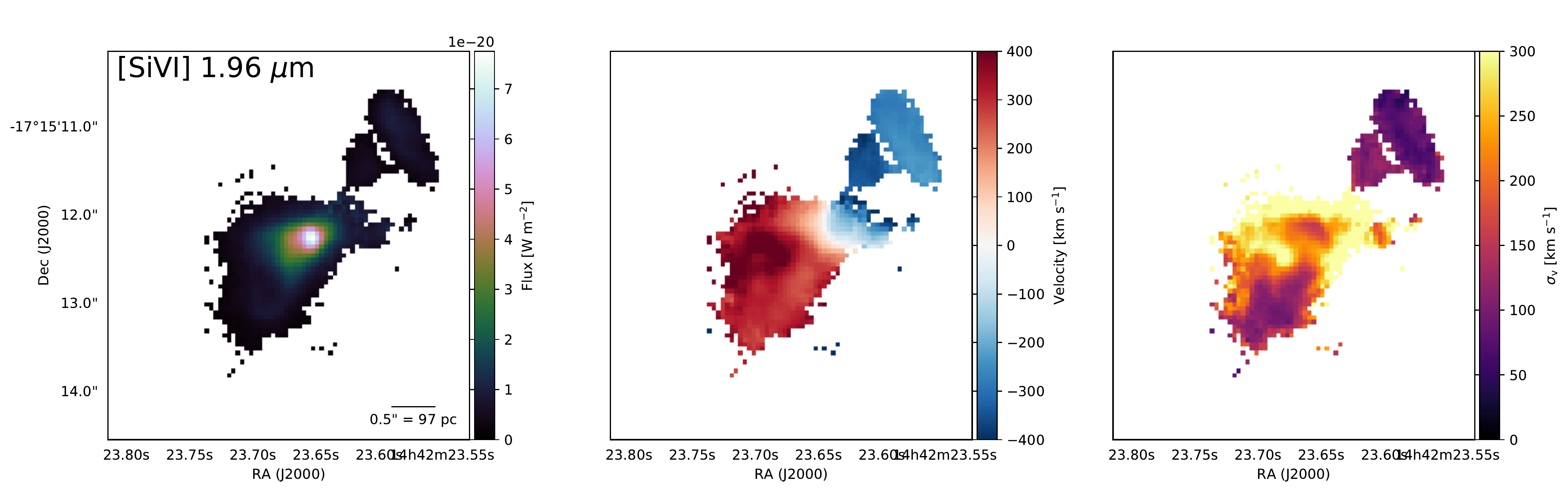}
\includegraphics[width=\columnwidth]{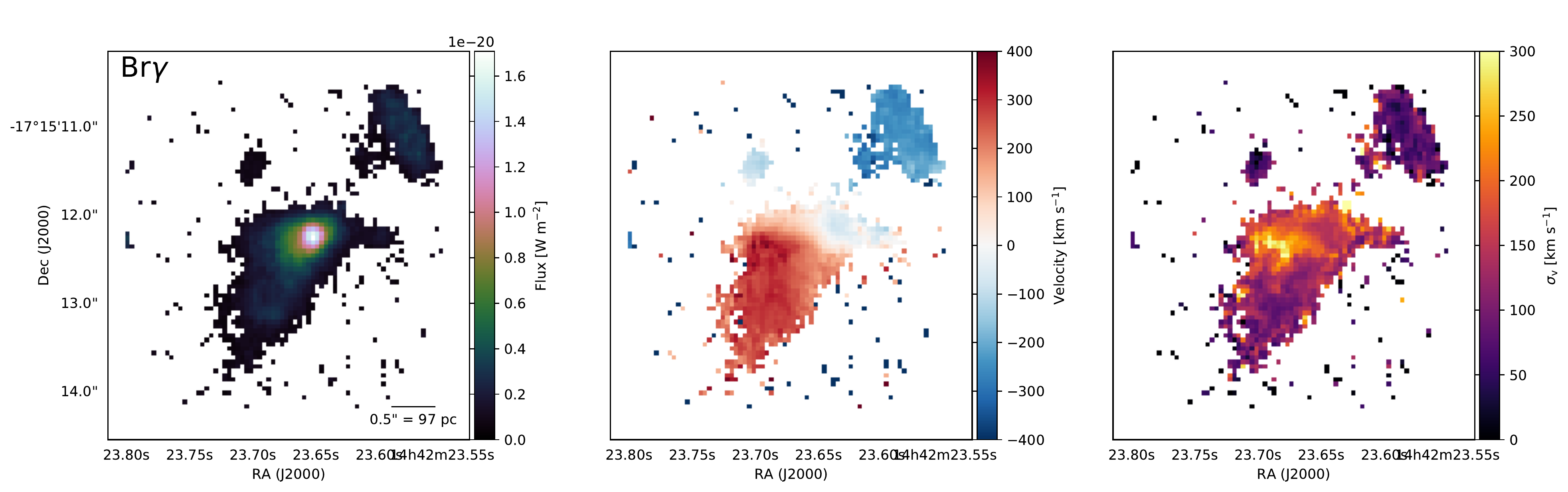}
\caption{\label{fig:ngc5728_sinfoni}\textit{Top:} [SiVI] 1.96 $\mu$m line flux, velocity, and velocity dispersion (left, middle, right panels) maps from VLT/SINFONI for NGC 5728. The biconical structure and high velocity strongly indicate and AGN driven wind. \textit{Bottom:} Same as above but for the Br$\gamma$ hydrogen recombination line at 2.16 $\mu$m. Br$\gamma$ shows the same geometry and velocity field as the higher ionization [SiVI] line emission. North is up and East is to the left in all panels.}
\end{figure}

\begin{figure}
\centering
\includegraphics[width=0.45\columnwidth]{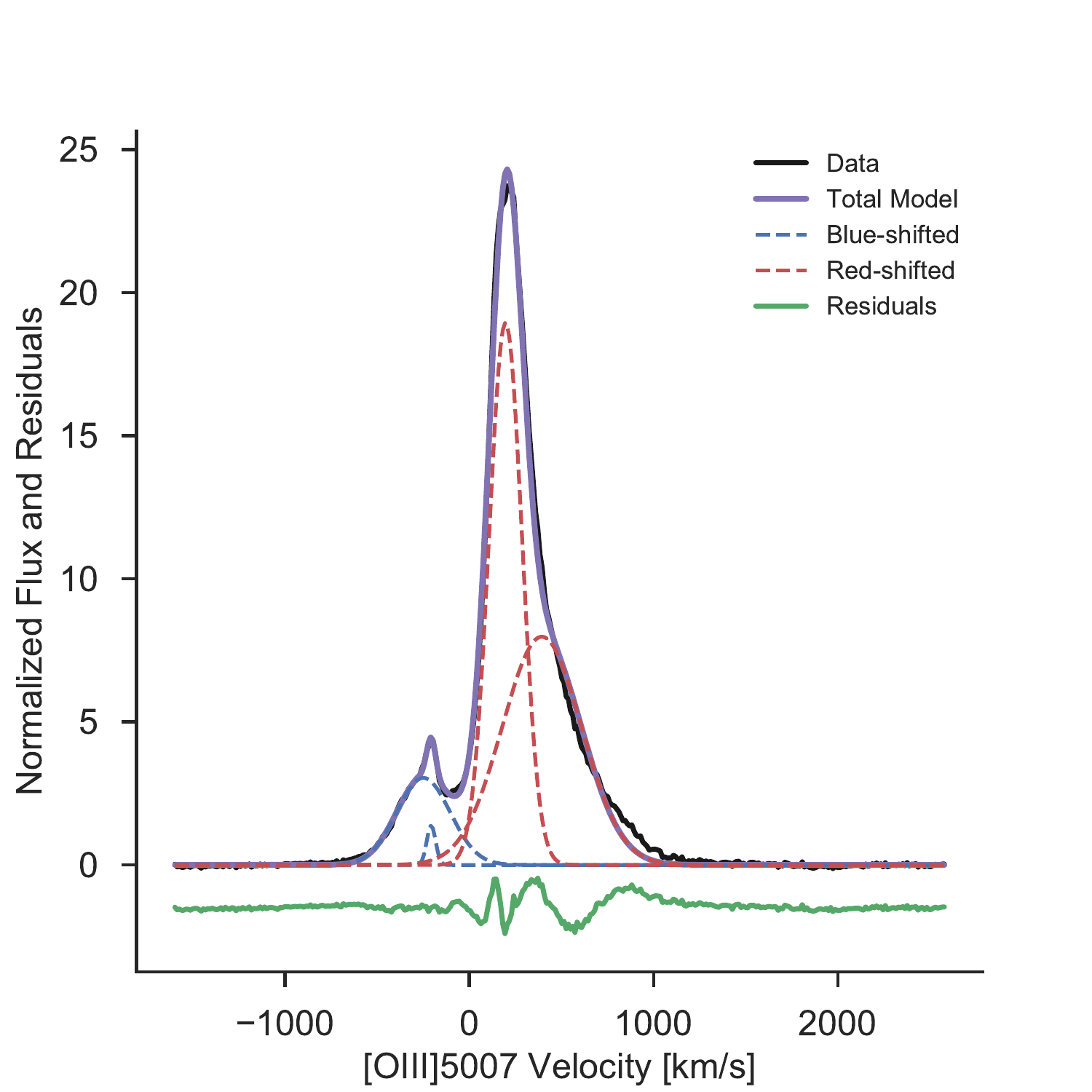}
\includegraphics[width=0.45\columnwidth]{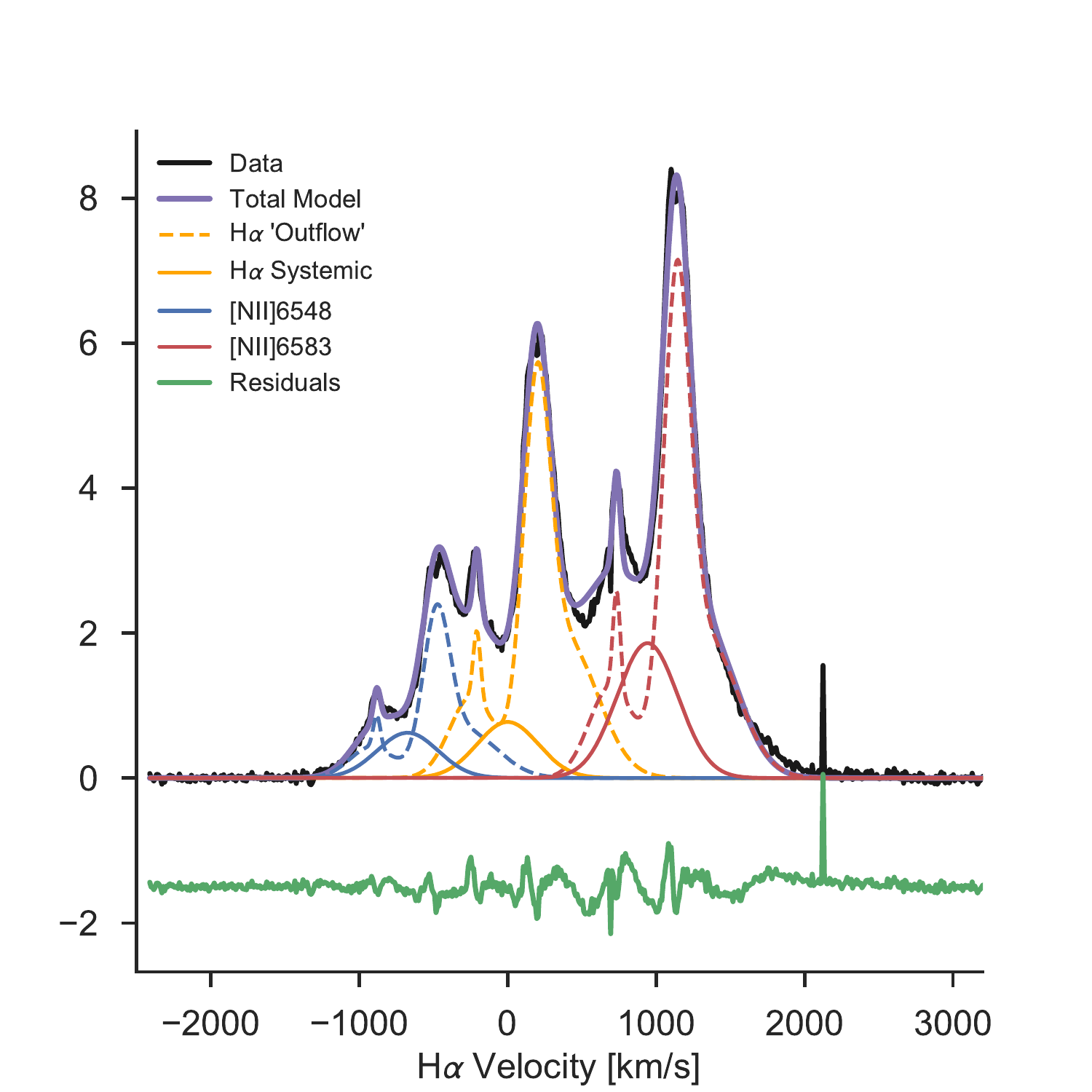}
\caption{\label{fig:ngc5728_x-shoot}\textit{Left panel:} X-Shooter [OIII]$\lambda$5007 emission profile for NGC 5728 (top; black) together with the best fit model (purple) containing four Gaussian components. Two components each were used for the blueshifted emission (blue dashed) and redshifted emission (red dashed). The green line displays the residuals after subtracting the best fit. \textit{Right panel:} X-Shooter \halpha+[NII] emission profile for NGC 5728 (top; black) together with the best fit model (purple). For each emission line, we included the same four Gaussian components that were needed for [OIII]$\lambda$5007 (dashed lines) and an extra systemic component (solid lines). The green line again displays the residuals. The x-axis indicates the velocity for the \halpha{} emission.}
\end{figure}

From this data set it is revealed that NGC 5728 contains a strong, spatially and spectrally resolved, wind. [SiVI] and Br$\gamma$ line emission maps from SINFONI (Fig~\ref{fig:ngc5728_sinfoni}) show the wind structure stretching from the SE to NW of the nucleus which matches the location and position angle of ionization cone seen in previous \textit{Hubble Space Telescope} narrow band \halpha+[NII] and [OIII] imaging \citep{Wilson:1993aa}. The SE half is redshifted while the NW half is blueshifted with each reaching up to a projected velocity of 400 km s$^{-1}$. The similarity in the flux and velocity maps of [SiVI] and Br$\gamma$ indicate the same process is driving the line emission for both species, likely AGN photoionization given the high ionization potential to produce [SiVI] (167 eV). Both the redshifted and blueshifted components of the outflow are seen in the X-Shooter spectrum as well. Fig.~\ref{fig:ngc5728_x-shoot} shows the [OIII]$\lambda$5007, and \halpha+[NII] spectral regions. We first fit the [OIII]$\lambda$5007 profile with four Gaussian components that reproduce the blueshifted broad bump and narrow peak and the redshifted narrow peak and wing. The velocities and velocity dispersion of these components are $(v, \sigma_{\rm v})=(-250\,\rm{km\,s^{-1}}, 145\rm{km\,s^{-1}})$ and $(-207\,\rm{km\,s^{-1}}, 24\,\rm{km\,s^{-1}})$ for the broad and narrow blueshifted components respectively and $(196\,\rm{km\,s^{-1}}, 89\,\rm{km\,s^{-1}})$ and $(393\,\rm{km\,s^{-1}}, 215\,\rm{km\,s^{-1}})$ for the redshifted narrow and wing components respectively. Using only these four components with fixed velocity and dispersion plus a 0~km~s$^{-1}$ velocity component with fixed velocity to account for a galactic disk component and fixing the line ratio of the [NII] to its theoretical value of 2.98, we can reliably reproduce the very complex \halpha+[NII] profile without the addition of any broad \halpha{} component. This indicates that NGC 5728 is in fact a Sy 2 with a strong ionized gas outflow, consistent with the high X-ray absorption.

\begin{figure}
\centering
\includegraphics[width=0.8\columnwidth]{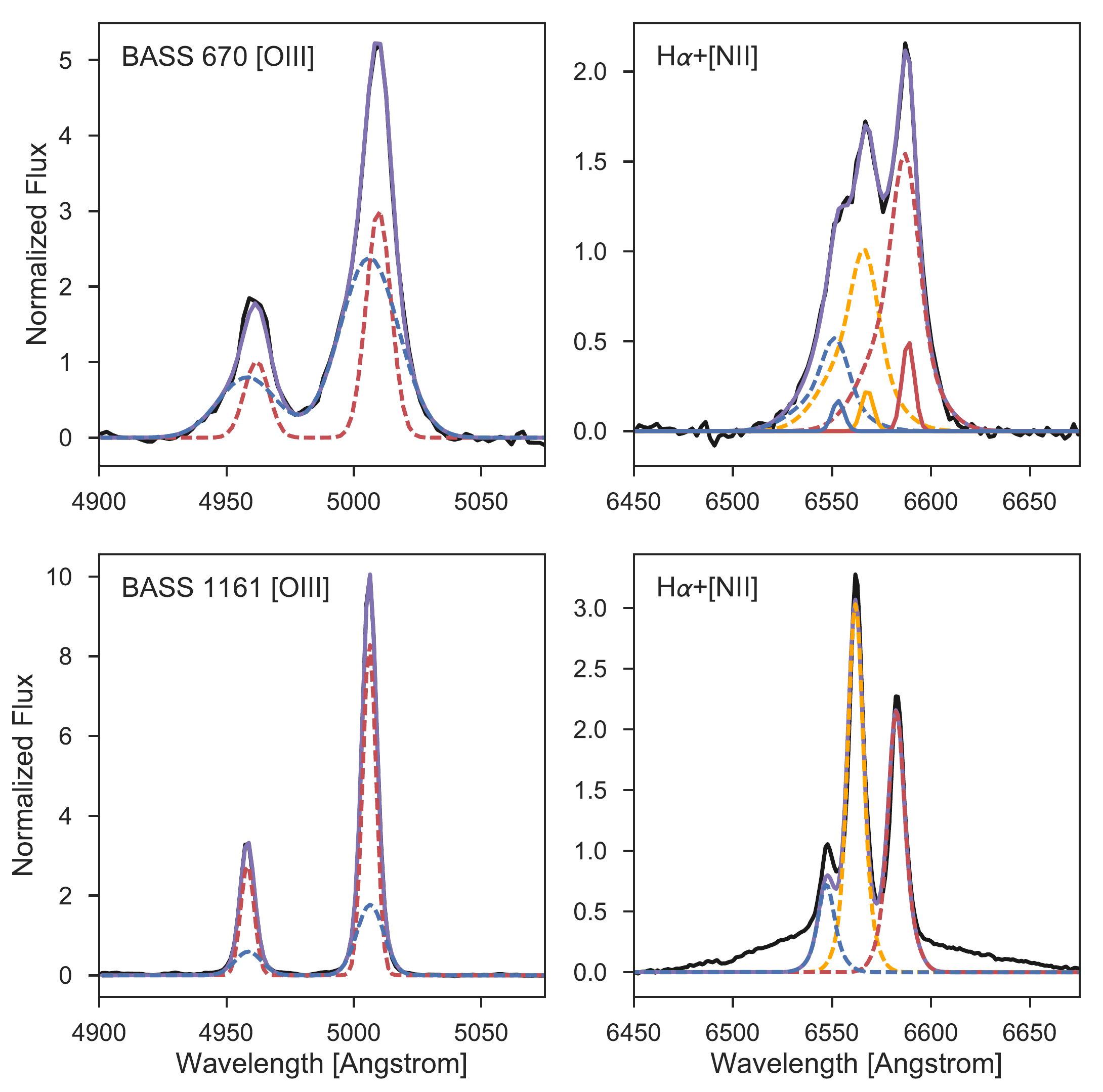}
\caption{\label{fig:example_fits}Example reanalysis for the [OIII] and \halpha+[NII] profiles of the Sy 1.9s. Top row shows the reanalysis of BASS 670 for which we were able to fit the \halpha+[NII] complex using the [OIII] Gaussian components plus an additional systemic component, thus indicating no compelling evidence for a BLR contribution to \halpha. The fit of the [OIII] doublet is shown in the left panel with the data in black, separate components blue and red dashed lines, and the total model in purple. The right panel shows the fit to the \halpha+[NII] complex with the combined [OIII] model shown as dashed blue, orange, and red lines and the additional systemic component as solid lines of the same color. The black and purple lines show the data and total fit, respectively. The bottom row is the same as above except for BASS 1161 for which there is strong evidence for the presence of a broad \halpha{} component.}
\end{figure}

\begin{figure}
\centering
\includegraphics[width=0.7\columnwidth]{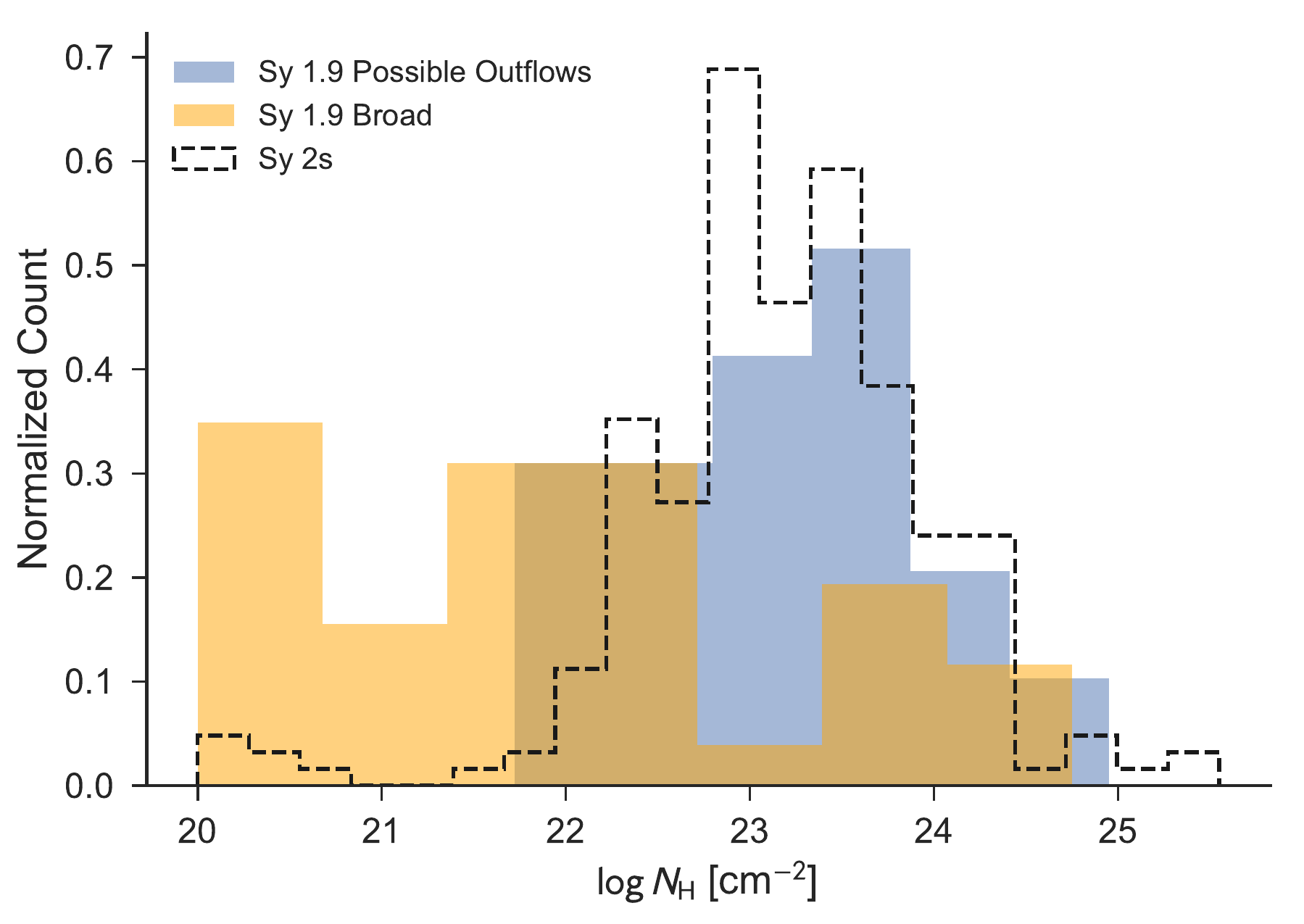}
\caption{\label{fig:outflow_nh}\nh{} distribution for Sy 1.9s that are possibly hosting outflows instead of a broad \halpha{} component (blue shaded histogram) compared to the \nh{} distribution for the remaining Sy 1.9s with evidence for a broad component (orange shaded histogram). The black dashed line indicates the \nh{} distribution for Sy 2s as shown in Fig.~\ref{fig:nh_dist}}
\end{figure}

Our analysis of NGC 5728 led us to re-evaluate the remaining BASS Sy 1.9s for the possibility that their broad \halpha{} components are actually part of an outflow. We repeated the methodology we used for NGC 5728, fitting the [OIII] profile with up to three Gaussian components, then fixing the velocity and velocity dispersions of these components to fit the \halpha+[NII] complex while also adding in a systemic component. Because of the lower spectral resolution, we simultaneously fit both the [OIII]$\lambda$4959 and [OIII]$\lambda$5007, fixing the line ratio to the theoretical value of 2.98. The final fit and residuals were inspected to determine whether there was evidence or not for an additional BLR component. In Fig.~{\ref{fig:example_fits}} we show examples for a source with no evidence for a BLR component (top row) and strong evidence for a BLR component. Out of 57 Sy 1.9s in our sample, we find that 32 show strong evidence for a broad \halpha{} component, 6 show weak evidence, and 18 show no evidence. For one source (BASS ID 929) we were not able to perform this analysis because [OIII]$\lambda5007$ is not detected. This indicates that up to 30\% of Sy 1.9s could in reality be Sy 2s with a strong outflow. Fig.~\ref{fig:outflow_nh} plots the \nh{} distribution for the Sy 1.9s with no evidence of a BLR component (blue shading) along with the remaining \nh{} distribution for those with strong and weak evidence (orange shading). We also show the Sy 2 \nh{} distribution from Fig.~\ref{fig:nh_dist} (dashed line) that matches well the distribution for Sy 1.9s with no BLR component. A K-S test on the two distributions indicates a 92\% probability for the null hypothesis that they are drawn from the same parent \nh{} population. 

In no way do we suggest that this analysis is conclusive and all 18 of the Sy 1.9s with no evidence of an underlying broad component have an outflow. Rather, with only moderate spectral resolution, it is possible to explain the \halpha+[NII] profile for them using only components present in the [OIII]$\lambda$5007 profile and an additional systemic component. Without at least higher spectral resolution data, we cannot make a conclusive statement for these objects. However, because the possibility remains that the broad \halpha{} component does not originate in the BLR, we choose to remove these 18 AGN from the rest of the analysis. The BASS collaboration is currently in the process of observing 75 BASS AGN with VLT/X-Shooter and the incidence of outflows will be discussed in future surveys.

\bibliographystyle{aasjournal}
\bibliography{/Users/ttshimiz/Dropbox/Research/my_bib}

\end{document}